\RequirePackage{fix-cm}
\documentclass[smallextended]{svjour3}       
\smartqed  
\usepackage{amsmath}
\usepackage{amssymb}
\usepackage{graphicx}
\usepackage{subfigure}
\usepackage[justification=centering]{caption}
\begin{document}

\title{Estimating the eigenstates of an unknown density operator without damaging it
}


\author{Jingliang Gao         \and
        Feng Cai 
}


\institute{Jingliang Gao \at
              State Key Laboratory of Integrated Services Networks, Xidian University, Xi'an 710071, China \\
              \email{gaojl0518@gmail.com}           
           \and
              Feng Cai \at
              xidian University
}

\date{Received: date / Accepted: date}

\maketitle

\begin{abstract}
Given $n$ qubits prepared according to the same unknown density operator $\rho$, we propose a nondestructive measuring method which approximately yields the eigenstates of $\rho$. It is shown that, for any plane which passes through the center point of the Bloch sphere, there exists corresponding projective measurement. By performing these measurements, we can scan the whole Bloch sphere like radar to search for the orientation of $\rho$ and determine the eigenstates. We show the convergency of the measurements in the limit $n\rightarrow\infty$. This result actually reveals a mathematical structure of the $n$-fold Hilbert space .
\end{abstract}

\section{Introduction}
\label{intro}
The quantum state estimation(or say quantum tomograph)\cite{Keyl2001,Keyl2006,Bagan2006,Gross2010,Renner2012,Bennett2006} is the task of inferring the density operator $\rho$ of a quantum system by appropriate measurements. Like a classical probability distribution
it cannot be measured on a single system, but can only be estimated on an ensemble sequence of identically prepared systems. Since $\rho$ can be spectral decomposed as $\rho=\sum_{i}p_{i}|e_{i}\rangle\langle e_{i}|$, the estimation of $\rho$ can be divided into two parts: the estimation of the spectrum $\{p_{i}\}$ and the estimation of the eigenstates $\{|e_{i}\rangle\}$. If we do not mind damaging the state, we can take measurements on each copy of $\rho$ and the estimations are simple. However, if we want the estimations are nondestructive, then the question becomes challenging. Instead of measuring each state individually, we need collective operations on all $n$ states simultaneously. For the spectrum, Keyl and Werner\cite{Keyl2001} proposed a scheme based on the irreducible representations of permutation group and showed that the Young frame is a good estimation. On the other hand, it is still unclear that what kind of mathematical structure can be used to estimate the eigenstates. In this paper, we focus on this problem and try to give a solution for the qubit systems.

\begin{figure}
\centerline{\includegraphics[scale=0.7]{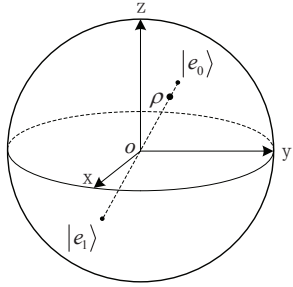}}
\caption{\label{fig:1} Bloch sphere}
\end{figure}

It is well known that a qubit system can be geometrically represented as a Bloch sphere\cite{Nielsen2000}. Suppose the spectrum decomposition of $\rho$ is $\rho=p|e_{0}\rangle\langle e_{0}|+(1-p)|e_{1}\rangle\langle e_{1}|$, then the eigenstates $|e_{0}\rangle, |e_{1}\rangle$ are determined by the axis $o\rho$ (see Fig.\ref{fig:1}). To estimate $|e_{0}\rangle$ and $|e_{1}\rangle$, we only need to know the orientation of the axis $o\rho$. Furthermore, the axis $o\rho$ is determined by the plane $zo\rho$ and the plane $yo\rho$, so the estimation of $|e_{0}\rangle, |e_{1}\rangle$ reduces to estimate the orientation of the plane $zo\rho$ and $yo\rho$. In the following, we will show that, for any plane which passes through the center point of the Bloch sphere, there exists a corresponding subspace in the $n$-fold Hilbert space. The projective measurement onto the subspace can distinguish whether $\rho$ lies in the plane or not. Successively performing these measurements, we can scan the Bloch sphere to locate the plane $zo\rho$ and $yo\rho$. It will be shown that this measuring scheme scarcely disturb the  state if $n$ is sufficiently large.
\section{Estimation of the eigenstates}
\label{sec:1}
Before giving the main results, we briefly introduce the strongly typical set\cite{Cover2012} which will be very useful for understanding our estimating method. We use the notation $x^{n}$ to denote a classical binary sequence $x_{0},x_{1},...,x_{n}$. $k$ is the number of $0$ in $x^{n}$. For a binary distribution $Q=(q, 1-q)$, the strongly typical set $A^{n}_{\varepsilon}(q)$ is defined as:
\begin{eqnarray}
A^{n}_{\varepsilon}(q)= \bigg \{ x^{n}: \left|\frac{k}{n}-q\right| \leq \frac{\varepsilon}{2} \bigg \}
\label{eq:1}
\end{eqnarray}
When $\varepsilon$ is small, $A^{n}_{\varepsilon}(q)$ is the collection of the sequences for which the number of $0$ is about $nq$, and the number of $1$ is about $n(1-q)$.

\begin{figure}
\centering
\subfigure[$: \|\rho^{\varphi}-I/2\|\leq\varepsilon$]{\includegraphics[scale=0.7]{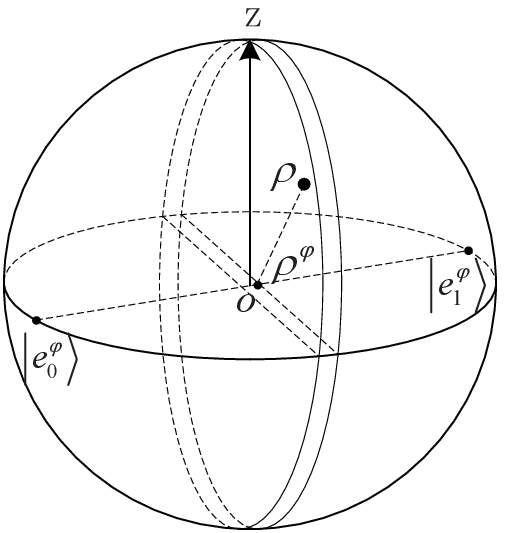}}
\hspace{0.1 in}
\subfigure[$: \|\rho^{\varphi}-I/2\|>\varepsilon$]{\includegraphics[scale=0.7]{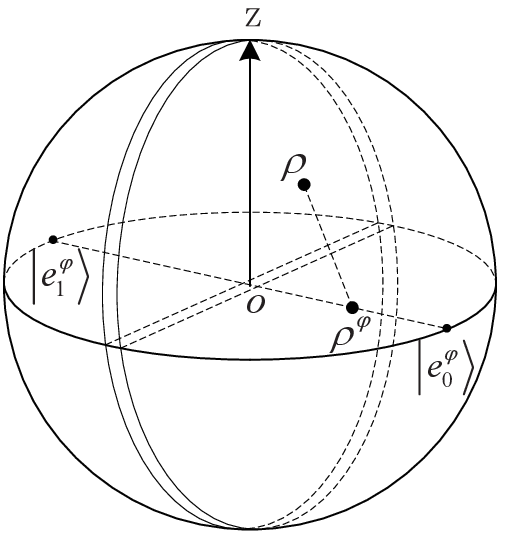}}
\caption{\label{fig:2} Geometric interpretation of Proposition1}
\end{figure}

Now we explain how to estimate the plane $zo\rho$. Define a basis $\{|e_{0}^{\varphi}\rangle, |e_{1}^{\varphi}\rangle \}$ by:
\begin{eqnarray}
|e_{0}^{\varphi}\rangle =\frac{1}{\sqrt{2}}\left( |0\rangle+e^{i\varphi}|1\rangle\right), |e_{1}^{\varphi}\rangle =\frac{1}{\sqrt{2}}\left( |0\rangle-e^{i\varphi}|1\rangle\right) \nonumber
\label{eq:2}
\end{eqnarray}
Clearly, this basis is corresponding to a diameter of the circle $xoy$. Let $\rho^{\varphi}=\langle e_{0}^{\varphi} |\rho|e_{0}^{\varphi}\rangle|e_{0}^{\varphi}\rangle\langle e_{0}^{\varphi}|+\langle e_{1}^{\varphi} |\rho|e_{1}^{\varphi}\rangle|e_{1}^{\varphi}\rangle\langle e_{1}^{\varphi}|$, then $\rho^{\varphi}$ corresponds to the projection of $\rho$ onto the basis $\{|e_{0}^{\varphi}\rangle, |e_{1}^{\varphi}\rangle\}$ (see Fig.\ref{fig:2}).\\
Construct a ``yes/no" measurement $\{M^{\varphi}_{yes}, M^{\varphi}_{no}\}$ by
\begin{eqnarray}
M^{\varphi}_{yes}&=&\displaystyle{\sum_{x^{n}\in A^{n}_{\varepsilon}(\frac{1}{2})}}|e_{x_{1}}^{\varphi}\rangle\langle e_{x_{1}}^{\varphi}| \otimes |e_{x_{2}}^{\varphi}\rangle\langle e_{x_{2}}^{\varphi}|\otimes ... |e_{x_{n}}^{\varphi}\rangle\langle e_{x_{n}}^{\varphi}| \nonumber \\
M^{\varphi}_{no}&=&I-M^{\varphi}_{yes}  \nonumber
\label{eq:3}
\end{eqnarray}
Perform this measurement on $\rho^{\otimes n}$, then we get the result ``Yes" with probability $tr(M^{\varphi}_{yes} \rho^{\otimes n})$.\\

\noindent \textbf{Proposition1}  \emph{For any $\varepsilon > 0$ and $\delta > 0$, when $n$ is sufficiently large, we have}\\
\\
(1-a) \emph{If $\|\rho^{\varphi}-I/2\|\leq \varepsilon$, then}
\begin{eqnarray}
tr(M^{\varphi}_{yes} \rho^{\otimes n}) \geq 1-\delta
\label{eqn:4}
\end{eqnarray}
\emph{where $\|\cdot\|$ is trace norm}.
\\

\noindent(1-b) \emph{If $\|\rho^{\varphi}-I/2\|> \varepsilon$, then}
\begin{eqnarray}
tr(M^{\varphi}_{yes} \rho^{\otimes n}) \leq\delta
\label{eq:5}
\end{eqnarray}
\\
(1-c) \emph{Let $F^{\varphi}$ be the entanglement fidelity}\cite{Nielsen2000} \emph{of the measurement, then}
\begin{eqnarray}
F^{\varphi} \geq 1-2\delta
\label{eq:6}
\end{eqnarray}\\

$I/2$ is the center $o$ of the Bloch sphere. Because $\rho^{\varphi}$ is the projection of $\rho$ onto the basis $\{|e_{0}^{\varphi}\rangle, |e_{1}^{\varphi}\rangle \}$, the condition $\|\rho^{\varphi}-I/2\|\leq \varepsilon$ implies that $\rho$ lies in a thin layer which takes $o$ as center and is perpendicular to the basis $\{|e_{0}^{\varphi}\rangle, |e_{1}^{\varphi}\rangle\}$ and is $\varepsilon$ thick(see Fig.2(a)). The proposition(1-a) reveals that if $\rho$ lies in the layer, the measurement will almost always give the result ``Yes". Proposition(1-b) reveals that if $\rho$ is outside the layer(see Fig.2(b)), the result will  almost always be ``No". So the measurement can distinguish whether $\rho$ lies in the thin layer or not. Since the parameter $\varepsilon$ can be as small as desired, the layer can be viwed as a plane. If the layer contains $\rho$, the layer approximate to the plane $zo\rho$! Furthermore, because $\delta$ can be arbitrary small, the measurement fidelity $F^{\varphi}$ is arbitrarily close to 1, which means that the measurement scarcely disturb $\rho^{\otimes n}$. Consequently, we can construct a sequence of measurements to find the plane $zo\rho$. We begin with the measurement $\{M^{\varphi=0}_{yes},M^{\varphi=0}_{no} \}$, if we get the result ``Yes", then $zo\rho$ is found; If ``No", perform the next measurement $\{M^{\varphi=\varepsilon}_{yes},M^{\varphi=\varepsilon}_{no} \}$. If we get the result ``Yes", then the work is done; If ``No", perform the next measurement $\{M^{\varphi=2\varepsilon}_{yes},M^{\varphi=2\varepsilon}_{no} \}$;  ...... As $\varphi$ increases, the layer rotates about the $z$-axis and scans the whole Bloch sphere gradually. Repeat the above procedures until for some $\varphi$ we get a ``Yes", then the orientation of $zo\rho$ is got.

A problem of this estimation scheme is that the fidelity decreases with the increasing times of measurement. However, this problem can be overcome if $n$ is large enough. Proposition(1-c) tell us that for any $\varepsilon$ and $\delta$, there exist a positive number $N(\varepsilon,\delta)$, if $n>N(\varepsilon,\delta)$, then $F^{\varphi}>1-\delta$.  Suppose we need $m$ times of measurement to find $zo\rho$ and the whole fidelity is $F$. Then we can reasonably assume that $F>1-f(m,\delta)$, where $f(m,\delta)$ is a continuous increasing function on $\delta$ and $m$ and $f(m,0)=0$, $f(1,\delta)=\delta$. For a given $\varepsilon$, $m$ is up to $\frac{\pi}{\varepsilon}$ because the measurement interval is $\varepsilon$. We can choose $\delta^{\ast}$ such that $f(\frac{\pi}{\varepsilon},\delta^{\ast})\approx0$, and choose $n$ such that $n>N(\varepsilon,\delta^{\ast})$. In this case, $F>1-f(m,\delta^{\ast})>1-f(\frac{\pi}{\varepsilon},\delta^{\ast})$, so $F\approx1$. In other word, when n is large enough, $F$ will be close to 1.

The plane $yo\rho$ can be estimated in the same way. We can define a basis $\{|e_{0}^{\theta}\rangle, |e_{1}^{\theta}\rangle \}$ as
\begin{eqnarray}
|e_{0}^{\theta}\rangle =\cos{\frac{\theta}{2}}|0\rangle+\sin{\frac{\theta}{2}}|1\rangle, \ \ |e_{1}^{\theta}\rangle =\sin{\frac{\theta}{2}}|0\rangle-\cos{\frac{\theta}{2}}|1\rangle \nonumber
\label{eq:7}
\end{eqnarray}
This basis corresponds to a diameter of the circle $xoz$.\\
Construct a measurement $\{M^{\theta}_{yes}, M^{\theta}_{no}\}$ by
\begin{eqnarray}
M^{\theta}_{yes}&=&\displaystyle{\sum_{x^{n}\in A^{n}_{\varepsilon}(\frac{1}{2})}}|e_{x_{1}}^{\theta}\rangle\langle e_{x_{1}}^{\theta}| \otimes |e_{x_{2}}^{\theta}\rangle\langle e_{x_{2}}^{\theta}|\otimes ... |e_{x_{n}}^{\theta}\rangle\langle e_{x_{n}}^{\theta}| \nonumber\\
M^{\theta}_{no}&=&I-M^{\theta}_{yes}  \nonumber
\label{eq:8}
\end{eqnarray}
Similarly, this measurement can tell us whether or not $\rho$ lies in the thin layer which is perpendicular to the basis $\{|e_{0}^{\theta}\rangle, |e_{1}^{\theta}\rangle \}$, and such a measurement scarcely disturb $\rho^{\otimes n}$. As $\theta$ increases, the layer rotates about the $y$-axis. Thus a measuring sequence can be constructed to find $yo\rho$.\\

\section{Proof of the Proposition}
Proving Proposition1 is quite a mathematical work and we need some knowledge about the method of types\cite{CK2011,Cover2012} which was developed by Csiszar and Korner. We still use $x^{n}$ to denote a binary sequence $x_{0},x_{1},...,x_{n}$ and $k$ is the number of $0$ in $x^{n}$.

The type $P_{\mathbf{x}}$ of $x^{n}$ is the relative proportion of $0$ and $1$, i.e. $P_{\mathbf{x}}(0)=k/n, \ \ P_{\mathbf{x}}(1)=1-k/n$. The collection of sequences of type $P$ is called the type class of $P$, denoted as $T(P)$, i.e. $T(P)=\left\{x^{n}:P_{\mathbf{x}}=P \right\}$. $|T(P)|$ is the number of elements in $T(P)$, then
\begin{eqnarray}
|T(P)| \leq 2^{nH(P)}
\label{eq:9}
\end{eqnarray}
where $H(\cdot)$ is the Shannon entropy function.\\
The strongly typical set $A^{n}_{\varepsilon}(q)$ can be stated as
\begin{eqnarray}
A^{n}_{\varepsilon}(q)=\bigcup\limits_{P:|P(0)-q|\leq \frac{\varepsilon}{2}}T(P)
\label{eq:10}
\end{eqnarray}
If we produce binary sequence according to the distribution $Q=(Q_{0},Q_{1}), Q_{0}=q$, then $A^{n}_{\varepsilon}(q)$ is a high probability set, i.e. for any fixed $\varepsilon > 0$ and $\delta > 0$, when n is large enough,
\begin{eqnarray}
Pr\{x^{n}\in A^{n}_{\varepsilon}(q)\}=\sum_{x^{n} \in A^{n}_{\varepsilon}(q)}Q(x^{n}) \geq 1-\delta
\label{eq:11}
\end{eqnarray}
where $Q(x^{n})=Q_{x_{1}}Q_{x_{2}}\cdots Q_{x_{n}}$ is the probability that $x^{n}$ occurs.
This property reveals that if we produce binary sequences according to the distribution $Q$, then the sequences are practically always in $A^{n}_{\varepsilon}(q)$. Furthermore, we give a stronger lemma.\\

\noindent \textbf{Lemma1:} \emph{Given $q$, for any distribution $Q^{\prime}=(q^{\prime}, 1-q^{\prime})$ which satisfies $|q^{\prime}-q|\leq \frac{\varepsilon}{2}$, $A^{n}_{\varepsilon}(q)$ is a high probability set, i.e. for any fixed $\varepsilon > 0$ and $\delta > 0$, when n is large enough,}
\begin{eqnarray}
\sum_{x^{n} \in A^{n}_{\varepsilon}(q)}Q^{\prime}(x^{n}) \geq 1-\delta
\label{eq:12}
\end{eqnarray}
\emph{where $Q^{\prime}(x^{n})=Q^{\prime}_{x_{1}}Q^{\prime}_{x_{2}}...Q^{\prime}_{x_{n}}$.}\\

This lemma tell us that if $Q^{\prime}$ is very close to $Q$, then $A^{n}_{\varepsilon}(q)$ is also a high probability set of $Q^{\prime}$.\\

\noindent \emph{Proof:} It is easy to show that $A^{n}_{\varepsilon}(q)$ contains a subset $A^{n}_{\varepsilon^{\prime}}(q^{\prime})$. Let $\varepsilon^{\prime}=\varepsilon-2|q^{\prime}-q|$. Given $|q^{\prime}-q|\leq \frac{\varepsilon}{2}$, it can be verified that if $\left|\frac{k}{n}-q^{\prime}\right| \leq \frac{\varepsilon^{\prime}}{2}$, then $\left|\frac{k}{n}-q\right| \leq \frac{\varepsilon}{2}$. So $A^{n}_{\varepsilon^{\prime}}(q^{\prime}) \subseteq A^{n}_{\varepsilon}(q)$. Thus we have
\begin{eqnarray}
\sum_{x^{n} \in A^{n}_{\varepsilon}(q)}Q^{\prime}(x^{n})\geq\sum_{x^{n} \in A^{n}_{\varepsilon^{\prime}}({q^{\prime}})}Q^{\prime}(x^{n})\geq 1-\delta
\label{eq:13}
\end{eqnarray}
The second inequality follows from (\ref{eq:11}).

Now we can prove the proposition1.

\noindent\emph{Proof:} 
\ Let $Q_{0}^{\varphi}=\langle e_{0}^{\varphi} |\rho|e_{0}^{\varphi}\rangle, Q^{\varphi}_{1}=\langle e_{1}^{\varphi} |\rho|e_{1}^{\varphi}\rangle$, then $\rho^{\varphi}=Q_{0}^{\varphi}|e_{0}^{\varphi}\rangle\langle e_{0}^{\varphi} |+Q^{\varphi}_{1}|e_{1}^{\varphi}\rangle\langle e_{1}^{\varphi} |$. Since $I/2$ can be decomposed as $I/2=\frac{1}{2}|e_{0}^{\varphi}\rangle\langle e_{0}^{\varphi} |+\frac{1}{2}|e_{1}^{\varphi}\rangle\langle e_{1}^{\varphi} |$, we have
\begin{eqnarray}
\|\rho^{\varphi}-I/2\|&=&tr\left|(Q_{0}^{\varphi}-\frac{1}{2})|e_{0}^{\varphi}\rangle\langle e_{0}^{\varphi} |+(Q^{\varphi}_{1}-\frac{1}{2})|e_{1}^{\varphi}\rangle\langle e_{1}^{\varphi} |\right| \nonumber \\
&=&|Q_{0}^{\varphi}-\frac{1}{2}|+|Q^{\varphi}_{1}-\frac{1}{2}| \nonumber \\
&=&|Q_{0}^{\varphi}-\frac{1}{2}|+|1-Q^{\varphi}_{0}-\frac{1}{2}| \nonumber \\
&=&2|Q_{0}^{\varphi}-\frac{1}{2}|   \nonumber
\label{eq:14}
\end{eqnarray}
First we prove (1-a). The condition $\|\rho^{\varphi}-I/2\|\leq \varepsilon$ is equivalent to $|Q_{0}^{\varphi}-\frac{1}{2}|\leq \frac{\varepsilon}{2}$. From Lemma1, we know that for the distribution $Q^{\varphi}=(Q_{0}^{\varphi},Q_{1}^{\varphi})$,
\begin{eqnarray}
\sum_{x^{n} \in A^{n}_{\varepsilon}(\frac{1}{2})}Q^{\varphi}(x^{n}) \geq 1-\delta
\label{eq:15}
\end{eqnarray}
where $Q^{\varphi}(x^{n})=Q_{x_{1}}^{\varphi}Q_{x_{2}}^{\varphi}...Q_{x_{n}}^{\varphi}$.
\begin{eqnarray}
tr(M_{yes}^{\varphi}\rho^{\otimes n})&=&\sum_{x^{n} \in A^{n}_{\varepsilon}(\frac{1}{2})}\left\langle e_{x_{1}}^{\varphi}e_{x_{2}}^{\varphi}...e_{x_{n}}^{\varphi}\right|\rho^{\otimes n}\left|e_{x_{1}}^{\varphi}e_{x_{2}}^{\varphi}...e_{x_{n}}^{\varphi}\right\rangle \nonumber\\
&=&\sum_{x^{n} \in A^{n}_{\varepsilon}(\frac{1}{2})}\prod_{i=1}^{n}\left\langle e_{x_{i}}^{\varphi}\right|\rho\left|e_{x_{i}}^{\varphi}\right\rangle \nonumber\\
&=&\sum_{x^{n} \in A^{n}_{\varepsilon}(\frac{1}{2})}Q_{x_{1}}^{\varphi}Q_{x_{2}}^{\varphi}...Q_{x_{n}}^{\varphi} \nonumber\\
&=&\sum_{x^{n} \in A^{n}_{\varepsilon}(\frac{1}{2})}Q^{\varphi}(x^{n}) \geq 1-\delta\nonumber
\label{eq:16}
\end{eqnarray}
So Proposition(1-a) is proven.

Next we give the proof of (1-b). For any type class $T(P)$ with $P=(P_{0},P_{1})$, we have
\begin{eqnarray}
&&\ \ \sum_{x^{n} \in T(P)}\left\langle e_{x_{1}}^{\varphi}e_{x_{2}}^{\varphi}...e_{x_{n}}^{\varphi}\right|\rho^{\otimes n}\left|e_{x_{1}}^{\varphi}e_{x_{2}}^{\varphi}...e_{x_{n}}^{\varphi}\right\rangle \nonumber \\
&&=\sum_{x^{n} \in T(P)}\prod_{i=1}^{n}\left\langle e_{x_{i}}^{\varphi}\right|\rho\left|e_{x_{i}}^{\varphi}\right\rangle \nonumber\\
&&=\sum_{x^{n} \in T(P)}\left\langle e_{0}^{\varphi}\right|\rho\left|e_{0}^{\varphi}\right\rangle^{nP_{0}}\left\langle e_{1}^{\varphi}\right|\rho\left|e_{1}^{\varphi}\right\rangle^{nP_{1}} \nonumber \\
&&=\sum_{x^{n} \in T(P)}2^{n[P_{0}\log{Q_{0}^{\varphi}}+P_{1}\log{Q_{1}^{\varphi}}]} \nonumber \\
&&=\sum_{x^{n} \in T(P)}2^{-n[H(P)+D(P||Q^{\varphi})]} \nonumber \\
&&\leq 2^{nH(P)}2^{-n[H(P)+D(P||Q^{\varphi})]} \nonumber \\
&&=2^{-nD(P||Q^{\varphi})} \nonumber
\label{eq:17}
\end{eqnarray}
where $D(P||Q^{\varphi})$ is the relative entropy\cite{Cover2012} between $P$ and $Q^{\varphi}$. The second equality holds because for any sequence $x^{n}$ with type $P$, the number of 0 is $nP_{0}$ and the number of 1 is $nP_{1}$. The inequality is due to (\ref{eq:9}).\\
Because $A_{\varepsilon}^{n}(\frac{1}{2})$ can be decomposed as (\ref{eq:10}), we have
\begin{eqnarray}
tr(M_{yes}^{\varphi}\rho^{\otimes n})&=&\sum_{x^{n} \in A^{n}_{\varepsilon}(\frac{1}{2})}\left\langle e_{x_{1}}^{\varphi}e_{x_{2}}^{\varphi}...e_{x_{n}}^{\varphi}\right|\rho^{\otimes n}\left|e_{x_{1}}^{\varphi}e_{x_{2}}^{\varphi}...e_{x_{n}}^{\varphi}\right\rangle \nonumber\\
&=&\sum_{P:|P_{0}-\frac{1}{2}|\leq \frac{\varepsilon}{2}} \sum_{x^{n}\in T(P)}\prod_{i=1}^{n}\left\langle e_{x_{i}}^{\varphi}\right|\rho\left|e_{x_{i}}^{\varphi}\right\rangle \nonumber\\
&\leq&\sum_{P:|P_{0}-\frac{1}{2}|\leq \frac{\varepsilon}{2}}2^{-nD(P||Q^{\varphi})} \nonumber \\
&\leq&\sum_{P:|P_{0}-\frac{1}{2}|\leq \frac{\varepsilon}{2}}2^{-n\min\limits_{P}{D(P||Q^{\varphi})}} \nonumber \\
&\leq& (n+1)\cdot2^{-n\min\limits_{P}{D(P||Q^{\varphi})}} \nonumber \\
&=&2^{-n[\min\limits_{P}{D(P||Q^{\varphi})}-\frac{\log{n+1}}{n}]}  \nonumber
\label{eq:18}
\end{eqnarray}
The third inequality holds because there are at most $n+1$ types for binary sequences of length $n$.\\
The condition $\|\rho^{\varphi}-I/2\|>\varepsilon$ is equivalent to $\left|Q_{0}^{\varphi}-\frac{1}{2}\right|>\frac{\varepsilon}{2}$. Combined with $\left|P_{0}-\frac{1}{2}\right|\leq \frac{\varepsilon}{2}$, we get $Q_{0}^{\varphi}\neq P_{0}$, so $D(P||Q^{\varphi})>0$. Thus when n is large enough, $2^{-n[\min{D(P||Q^{\varphi})}-\frac{\log{n+1}}{n}]}$ can be arbitrarily small, i.e. for any $\delta>0$, $tr(M_{yes}^{\varphi}\rho^{\otimes n})<\delta$.

The proof of (1-c) is very simple. According to\cite{Nielsen2000}, for any quantum operation $\mathcal{E}(\rho)=\sum_{k}E_{k}\rho E^{\dag}_{k}$, the entanglement fidelity $F(\rho, \mathcal{E})=\sum_{k}|tr(E_{k}\rho)|^{2}$. So for the measurement $\{M^{\varphi}_{yes}, M^{\varphi}_{no}\}$,
\begin{eqnarray}
F^{\varphi}&=&\left|tr(M^{\varphi}_{yes}\rho^{\otimes n})\right|^{2}+\left|tr(M^{\varphi}_{no}\rho^{\otimes n})\right|^{2} \nonumber \\
&\geq&\left|1-\delta\right|^{2} \geq 1-2\delta    \nonumber
\label{eq:19}
\end{eqnarray}
This complete our proof.\\
\section{Conclusion}
In this paper, we proposed a method for estimating the eigenstates of $\rho$. We show that the orientation of $zo\rho$ and $yo\rho$ can be found via a sequence of ``yes/no" measurements and then the eigenstates can be determined.


%

%

\begin{acknowledgements}
This work is supported by the National Natural Science Foundation of China Grant No.61271174, No.61372076 and No.61301178.
\end{acknowledgements}



\end{document}